\documentclass[10pt,conference]{IEEEtran}
\IEEEoverridecommandlockouts
\usepackage{cite}
\usepackage{amsmath,amssymb,amsfonts}

\usepackage[Procedure,ruled]{algorithm}
\usepackage[noend]{algpseudocode}

\usepackage{graphicx}
\usepackage{textcomp}
\usepackage{xcolor}
\usepackage{todonotes}
\usepackage{tikz}
\usepackage{amssymb}
\usepackage{pifont}

\usepackage{url}
\usepackage{breakurl}
\usepackage[breaklinks]{hyperref}
\usepackage{listings,xcolor}

\definecolor{dkgreen}{rgb}{0,.6,0}
\definecolor{dkblue}{rgb}{0,0,.6}
\definecolor{dkyellow}{cmyk}{0,0,.8,.3}

\newlength\myindent
\author{
    \IEEEauthorblockN{Chaofan Shou, İsmet Burak Kadron, Qi Su, and Tevfik Bultan}
    \IEEEauthorblockA{University of California, Santa Barbara
    \\\{shou, kadron, qisu, bultan\}@cs.ucsb.edu}

}
\begin{document}

\title{CorbFuzz: Checking Browser Security Policies with Fuzzing\\
\thanks{This material is based on research supported by NSF under Grants CCF-1901098 and CCF-1817242.}
}

\maketitle

\begin{abstract}
Browsers use security policies to block malicious behaviors. Cross-Origin Read Blocking (CORB) is a browser security policy for preventing side-channel attacks such as Spectre. We propose a web browser security policy fuzzer called CorbFuzz for checking CORB and similar policies. In implementing a security policy, the browser only has access to HTTP requests and responses, and takes policy actions based solely on those interactions. In checking the browser security policies, CorbFuzz uses a policy oracle that tracks the web application behavior and infers the desired policy action based on the web application state. By comparing the policy oracle with the browser behavior, CorbFuzz detects weaknesses in browser security policies. CorbFuzz checks the web browser policy by fuzzing a set of web applications where the state-related queries are symbolically evaluated for increased coverage and automation. CorbFuzz collects type information from database queries and branch conditions in order to prevent the generation of inconsistent data values during fuzzing. We evaluated CorbFuzz on CORB implementations of Chromium and Webkit, and Opaque Response Blocking (ORB) policy implementation of Firefox using web applications collected from GitHub. We found three classes of weaknesses in Chromium's implementation of CORB.

\end{abstract}

\section{Introduction}

Web browsers allow users to various things, such as streaming videos or accessing bank accounts. A malicious website should not be able to access sensitive information about a web application user, for example a bank account page. Unfortunately, due to vulnerabilities like cross-site script inclusion~\cite{hailperin_cross-site_nodate}, cross-site scripting~\cite{grossman2007xss}, Spectre~\cite{kocher2019spectre}, and Meltdown~\cite{Meltdown}, malicious websites can access sensitive information that they should not have access to. Because of the aforementioned threats, browsers have adopted an increasing number of security policies like Cross-Origin Read Blocking (CORB) policy~\cite{corb} that they use to protect sensitive data. The goal of the CORB policy is to prevent cross-origin access to confidential data. 

In order to determine if a behavior is malicious or not, a browser security policy has to infer properties about the web application that is being used. Yet, given that a browser does not have access to web applications' internal state, nor its codebase, it cannot precisely determine the properties of the web applications. Instead, security policy implementations use the information browsers have access to, like HTTP responses and requests, to infer properties of web applications and decide to take a policy action according to those properties. 

In this paper, we focus on CORB as a browser security policy because it is one of the most important policies for protecting cross-origin resources. CORB aims to identify and block all cross-origin loads of confidential response content. However, browsers can not determine whether a specific response is confidential without inspecting the state of the web application. Since the browser cannot do that, the CORB policy implementations examine the responses instead and use information inside responses and heuristics that reflect the expected behavior to determine whether the content is confidential.


These heuristic approaches need to be tested comprehensively in order to look for scenarios where they fail to protect sensitive information. A fully automated testing approach would enable browser security policy developers to identify weaknesses in existing policies and to quickly evaluate policy modifications. 



We developed a fuzzing technique to check browser security policies. 
Given a browser and a security policy, we use a set of open-source web applications to look for weaknesses in the security policy implementation of that browser. We use the open-source web applications as fuzzing targets, and our fuzzer creates requests for each of them, intending to achieve as much coverage as possible. By exploring a variety of web applications and covering as many behaviors as possible for each web application, our fuzzer tests a large set of scenarios for the browser security policy implementation. 


In order to identify weaknesses in the browser policy implementation, we define a reference implementation of the security policy by tracking the web application states and utilize it as an oracle. The oracle is more accurate than the browser policy implementation since during fuzzing, the oracle has access to all internal information of the web application and properties of each response. Our fuzzer compares the decisions made by the oracle to the security policy implementation of the browser and reports any differences, which correspond to a weakness in the browser security policy implementation.

Most web applications typically access session data, cookies, and data store~\cite{web_two_decades}. These web applications are called data-dependent. Fuzzing a data-dependent web application requires manually setting up these data sources (e.g., populating a database~\cite{alshahwan2011automated,biagiola2017search}). However, given that we need to use a set of web applications during fuzzing, it is not practical to manually set up the data store and session values for each web application. Thus, we propose a runtime for data-dependent web applications that enables us to automate the process.  
Instead of manually setting up data sources for all fuzzing targets, our runtime automatically synthesizes data store, sessions and cookies. This approach not only removes the requirement for manually setting up an environment for a web application, but also allows our fuzzer to easily mutate the data store, leading to higher coverage.

The runtime we propose generates SMT constraints for database queries and sessions or cookies usage. The SMT constraints for a database query encode the SQL statement, and we use an SMT solver to generate data values consistent with the query. The SMT constraints generated for sessions and cookies are used to check the feasibility of execution paths of the web application.

Using this approach, we have implemented a fuzzer focusing on the CORB policy, which we call CorbFuzz. While CorbFuzz is optimized for CORB analysis, it can be easily extended to support other policies by defining corresponding oracles. Additionally, for our prototype, we restrict our scope to PHP applications. Our approach can be extended to support {web applications developed in different programming languages, including Python or NodeJS,} by providing simple instrumentation for the target language as discussed in Section~\ref{sec:data-synthesis}. 

We evaluate the implementation of CORB policy in both Chromium and Webkit. CorbFuzz did not find any policy violations in Webkit and shows that CORB implementation in Webkit is robust. In Chromium, on the other hand, CorbFuzz identifies three types of code patterns that can enable attackers to bypass CORB protection. Furthermore, we modified CorbFuzz to check a sibling policy by Firefox called Opaque Response Blocking (ORB). 

In this paper we present the following research contributions:
\begin{itemize}
\item {\em Browser Policy Fuzzer:} We propose a new fuzzer, CorbFuzz, for checking browser security policies. CorbFuzz is guided by web application code coverage and uses a policy oracle to identify weaknesses in browser security policies. It is fully automated and can be easily applied after each change in policy implementation. 
\item {\em Data Synthesis:} To tackle fuzzing environment setup for data-dependent applications, we propose a runtime that synthesizes and mutates the data when required. Our data synthesis approach uses SMT encoding and constraint solving to ensure consistency of data generated for database queries and sessions or cookies usage.
\item {\em Empirical Evaluation:} We used CorbFuzz to check the CORB implementation of Chromium and Webkit. We also checked a sibling policy ORB for Firefox.
We fuzzed these policies using responses of PHP web applications that we obtained from GitHub. 
Using CorbFuzz, we discovered three code patterns that expose weaknesses in the CORB implementation of Chromium. One of these code patterns has been previously documented, and the Chromium team patched the policy weakness caused by another code pattern we discovered after our report. 
\end{itemize}

The paper is structured as follows. 
In Section~\ref{sec:background}, we present the background on browser precautions. 
{
In Section~\ref{sec:methodology}, we present our fuzzing framework.
In Section~\ref{sec:data-synthesis}, we discuss how we synthesize the data and bypass authentication for web applications.}
In Section~\ref{sec:evaluation}, we evaluate CorbFuzz and describe the detected CORB weaknesses by our tool.
In Section~\ref{sec:related-work}, we present the related work.
In Section~\ref{sec:conclusion}, we conclude the paper.

\section{Background}
\label{sec:background}
In this section, we provide the background information on Site Isolation and Cross-Origin Read Blocking policy.



\subsection{Site Isolation and Information Leakage}
Browser information leakage has gained increasing exposure in the last few years. According to the Same-Origin Policy (SOP)~\cite{203852}, one of the fundamental rules in browsers, documents from different origins cannot interact with each other. However, many exploits have been discovered to conduct cross-origin content leak~\cite{gulmezoglu_zankl_eisenbarth_sunar_2017,jana_shmatikov_2012,kim_lee_kim_2016,lee_kim_kim_kim_2014,spreitzer_griesmayr_korak_mangard_2016,karami_ilia_polakis_2021}. Additionally, the discovery of cache-related side-channel vulnerabilities like Spectre\cite{Spectre} and Meltdown\cite{Meltdown} worsen the information leaks.

Site Isolation policy~\cite{siteisolation,siteisolationchromium} has been proposed to counter cross-origin content leaks. Such a policy is also known as ``one site per process'' policy. Namely, a browser should ensure that documents from different origins are rendered and executed in their own respective sandbox. Such an effort reduces the chance of success of cache side-channel attacks and makes most cross-origin information leakage vulnerabilities in browsers no longer exploitable. 



\subsection{Cross-Origin Read Blocking}
\label{corb-background}
While Site Isolation policy removes the possibility of documents in different origins interacting with each other directly, there are still ways to inject documents from different origins via interfaces provided by browsers. A possible approach is to include the documents from different origins as resources required by the webpage. Some examples have been provided below, for which the first line is to load an endpoint as an image, and the second line is to load it as a script.

{\footnotesize
\begin{verbatim}
<img src="//a.com/secret" />
<script src="//a.com/secret"></script>
\end{verbatim}
}





In addition, other browser JavaScript interfaces could be used to pass partial sensitive information from one origin to another. A famous example is CVE-2020-6442 \cite{noauthor_cve_nodate}. The vulnerability is that by loading two cross-origin documents into the cache, it is possible to calculate the difference of sizes between two documents by calculating the increase in the size of the cache. The size leakage technique could be easily exploited to deduce the preference and the visiting history of users. 

All these interactions make Site Isolation policy no longer effective. While blocking all cross-origin requests could solve the issue, existing websites legitimately utilizing cross-origin resources would similarly be affected by such an approach. Thus, Cross-Origin Read Blocking (CORB) policy has been proposed. It aims to prevent HTTP responses from being loaded into contexts at different origins if the information is deemed confidential. The authors have claimed that this could effectively reduce potential dubious cross-origin resource fetches. Previous examples of XSSI attacks or the CVE-2020-6442 vulnerability become ineffective when a browser implements the CORB policy~\cite{noauthor_1013906_nodate}. 

A simplified version of the CORB policy implementation in Chromium is shown in Procedure~1. This code is executed as soon as a response is received by the browser. It performs a few initial checks, including whether the scheme is HTTP(S). If these checks are not violated, the response is allowed to be loaded into a context in a different origin (i.e., not blocked). The procedure returns NULL if the response is blocked. 

The CORB policy authors defined a set of response MIME types likely related to secrets, namely protected MIME types. The response having Content Type header value as a protected MIME type is blocked. For instance, responses with Content Type headers related to images would not be blocked, yet responses with Content Type headers related to JSON are blocked because web developers commonly use JSON serialized responses to conduct communication between frontend and backend. 

Chromium team took a different approach to implement CORB. Instead of strictly following the policy documented at W3C\cite{noauthor_fetch_nodate}, the team added extra measures to confirm the MIME types by inspecting the response content~\cite{noauthor_mime_nodate}. This measure is known as ``confirmation sniffing''. They claimed that this could effectively reduce false positives (i.e., reduce the cases when a legitimate response is blocked), thus increasing the compatibility of Chromium with more web applications~\cite{corb}. For instance, as seen in Lines~6 and 7 in Procedure~1, if the response MIME type is related to JSON, which is in the protected MIME type list, but the content in the response is an image, not a JSON, then Chromium follows the property of the content and does not block. On the other hand, Webkit strictly follows the policy and blocks the response since it does not have such a measure~\cite{webkitcorb}.

\label{proc:corb-imp}
\begin{algorithm}
\begin{algorithmic}[1]
    \Procedure{CorbCheck}{Response}
    \If{Response.Scheme $\notin$ \{HTTP, HTTPS\}}
        \State \Return Response
    \EndIf
    \State mime $\leftarrow$ Response.ContentType 
    \If{mime $\in$ ProtectedMimeTypes}
        \If{mime $\in$ JSON $\wedge\neg$ IsJSON(Response)}
            \State \Return Response
        \EndIf
        \If{mime $\in$ XML $\wedge\neg$ IsXML(Response)}
            \State \Return Response
        \EndIf
        \Else
            \State \Return NULL
    \EndIf
    \State \Return Response
    \EndProcedure
\end{algorithmic}
    \caption{Partial CORB Implementation in Chromium}

\end{algorithm}

\section{Browser Policy Fuzzing}
\label{sec:methodology}

\begin{figure} [t] \small
    \centering
    \includegraphics[width = 80mm]{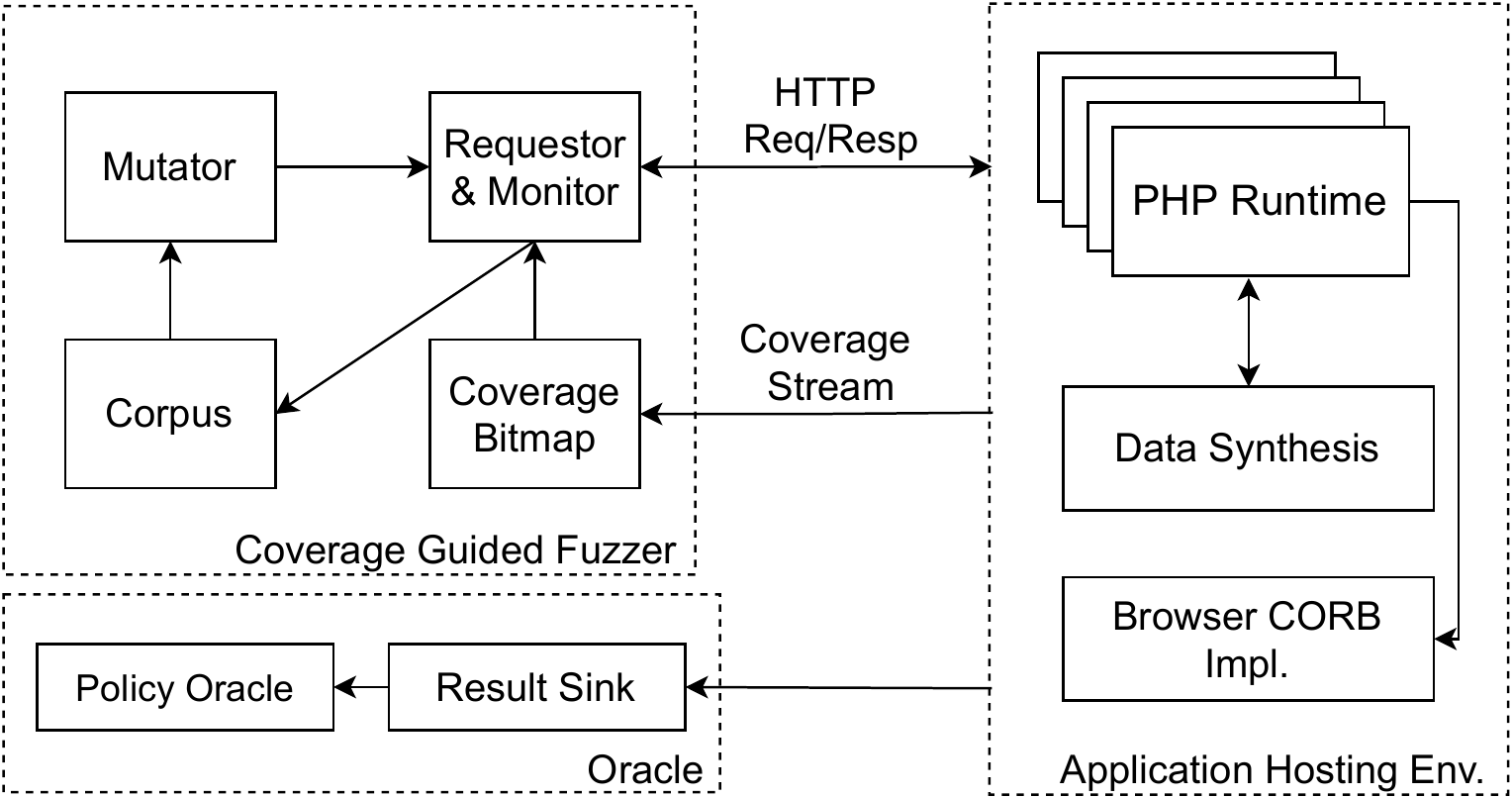}
    \caption{CorbFuzz Architecture}
    \label{fig: arch corbfuzz}
\end{figure}

In this section we present CorbFuzz, which is a fuzzing technique for
checking browser security policies. 
%


\label{proc:corbfuzz-imp}
\begin{algorithm}
\begin{algorithmic}[1]

\Procedure{CorbFuzz}{TestBench}

    \For{WebApplication $\in$ TestBench}
        \State \textsc{CorbFuzzTestOne}(WebApplication)
    \EndFor
\EndProcedure
\Procedure{CorbFuzzTestOne}{WebApplication}
    \State $P \leftarrow \text{DataSynthesis}(\text{WebApplication})$
    \State CovBitMap $\leftarrow$ BitMap() 
    \State ResultSink $\leftarrow$ HashTable() 
    \State Corpus $\leftarrow$ List[Pair]() 
    \State NewURL $\leftarrow$ List()
    \State Visited $\leftarrow$ Set()

    \For{$\neg$ShouldTerminate()}
        \State U, Seed $\leftarrow$ (Corpus $\cup$ (NewURL - Visited)).Pop()
        \State U, Seed $\leftarrow$ Mutate(U, Seed)
        \State Visited $\overset{+}{\leftarrow}$ U
        \State Metrics, R, M $\leftarrow$ $P$(U, Seed) 
        \State NewURL $\overset{+}{\leftarrow}$ ExtractLinks()
        \If{IsNewCoverage(CovBitMap, Metrics)}
            \State Corpus $\overset{+}{\leftarrow}$ U, Seed
            \State CovBitMap $\overset{+}{\leftarrow}$ Metrics
        \EndIf
                \If{IsUniqueResponse()}

                    \State ResultSink $\overset{+}{\leftarrow}$ (R, M)
                    \EndIf

    \EndFor
    \State \text{RunOracle}(\text{ResultSink})
\EndProcedure

\end{algorithmic}
    \caption{CorbFuzz Algorithm} 
\end{algorithm}

\subsection{Fuzzing Algorithm}

We present the architecture of CorbFuzz in Figure~\ref{fig: arch corbfuzz} and its algorithm in Procedure~2. CorbFuzz is a distributed and multi-threaded fuzzer that loops over all given web applications and calls \textsc{CorbFuzzTestOne}, which conducts coverage-guided fuzzing for the given web application individually. 

Initially, \textsc{CorbFuzzTestOne} creates multiple instances of the application runtime {instrumented with data synthesis discussed in Section-\ref{sec:data-synthesis}}. We define the runtime to be a function $P: (\text{URL}, \text{Seed}) \to \{\text{Metrics}, \text{R}, \text{M}\}$, where {$\text{Seed}$ is an identifier mapping to a state of the web application (i.e., database, cookies, and sessions)}, $\text{Metrics}$ represents the coverage metrics, $\text{R}$ represents the resource queried by the web application, and $\text{M}$ maps each CORB implementation to its decision on whether to block the response. Analogous to a pipeline, the HTTP requests are first passed to the runtime hosting web application and HTTP responses generated are then served as inputs for different CORB implementations. 

In \textsc{CorbFuzzTestOne}, a bitmap is created so as to record the coverage (Procedure 2, Line 6).  \textsc{CorbFuzzTestOne} additionally declares a result sink (Line 7) for storing the information required by the oracle.
The details of the oracle are elaborated in the following sections. 
A corpus (Line 8) is also defined as a list of pairs, where each pair contains the URL of the request and the seed. 
Additionally, a list is declared (Line 9) for storing the URLs extracted from the HTTP response (e.g., href values and API calls). 
During fuzzing in \textsc{CorbFuzzTestOne}, it randomly selects an input from either corpus or unvisited links extracted (Line 9). The input contains a request URL and potentially a seed mapping to a state. Then, \textsc{CorbFuzzTestOne} mutates and sends the corresponding HTTP request and seed to the runtime (Lines 10, 12). If the input leads to increased coverage, it is added to the corpus (Lines 14-15). 

After the fuzzing terminates (i.e., ShouldTerminate() returns true), the oracle aggregates the information in result sink and provides a decision for each HTTP response. These decisions are compared with the browser decisions to identify potential weaknesses.

\subsection{Policy Oracle}
To define a policy oracle (i.e., RunOracle function in Procedure 2, Line 22), we need to categorize the response as confidential or non-confidential by evaluating resource accesses. We limit the scope of resources to be only provided by the database for this work. We use a method similar to Pellegrino et al.~\cite{pellegrino2017deemon} which deduces confidentiality of a resource by observing resource access frequency. After fuzzing terminates for each web application, we aggregate and count the number of resources accessed by each database query executed while handling each request. In our implementation, we use the average number of accesses as our threshold. If any query uses resources that have a frequency below the threshold, the oracle infers that the query is accessing a confidential resource, of which the response should be blocked, and checks whether the CORB implementation blocked it. 

The granularity of the resource impacts the result of the oracle. For example, if each resource is considered as a table, oracle is more likely to decide to block the response than if each individual resource is considered as a row in the table. Hence, using a coarse-grain resource definition is more likely to produce false positives. We designed two types of oracles with different resource granularity. To reduce false positives, one oracle considers each unique row (i.e., query constraints) to be a resource, and to reduce false negatives, the other oracle considers each table to be a resource.    







\subsection{Coverage Metrics}
In \textsc{CorbFuzzTestOne}, rather than focusing on test coverage for the CORB function of the browser implementation, fuzzing is guided by test coverage for the given web application. While the coverage information of CORB function may enhance the fuzzing in regards to efficiency, it would not be useful since CORB function is a small piece of code in both WebKit and Chromium. Thus, it is easy to achieve high test coverage for the CORB function while focusing on achieving high test coverage for the given web application. Additionally, we are evaluating the policy for different code patterns. Focusing on what CORB is able to handle would not lead to identifications of potential weaknesses in the implementation. 

 
\section{Data Synthesis}
\label{sec:data-synthesis}

In this section, we discuss our data synthesis techniques that enable us to handle data-dependent web applications automatically during fuzzing (i.e., during the execution of \textsc{CorbFuzzTestOne}), without the need for manual set up of fuzzing targets. Instead of querying the database, the data synthesis approach translates the query to constraints and generates the respective data. Additionally, the data synthesis approach generates results for comparisons involving sessions or cookies so as to achieve higher test coverage and bypass authentication. 

The data synthesis workflow depends on a seed that is generated and tracked by \textsc{CorbFuzzTestOne}, first mentioned in Procedure~2, Line~13. A seed is a 32-bit integer sampled from a uniform distribution over $[0, 2^{32}-1]$. There is a bijection between the seed and the state of the database and a weak bijection between the seed and the cookies or sessions.




\subsection{Query Constraint Extraction}
\label{subsec:query-constraint-extraction}
We first discuss the handling of database queries. 
The results generated for a specific database query are constrained by three measures: row count, table architecture, and constraint that describes the resulting rows and columns from the query. Most open-source web applications either do not include a table schema or require laborious work to set up the tables. Thus, we assume that the table schema is not given, and the generation of the database query result is run without the knowledge of the table architecture. For these, we respectively define three functions: \textsc{MaxRow}, \textsc{Fields}, \textsc{Constraint}. The input of all these functions is a relational algebra expression translated from the query. 

\textsc{MaxRow} provides an estimation of the maximum rows of the query result. It is implemented by considering the set operators and \textit{LIMIT}.  

To reconstruct the table schema, the data synthesis approach learns from the query by observing the field names used inside it. We define the \textsc{Fields} function, which produces a set of pairs representing fields returned by the query. The first part of the pair indicates the table name, and the second part is the name of the field. The function is implemented by tracking the rename, projection, and select operators. In the case that a wildcard projection (i.e., asterisk) is used, the function only returns the fields used throughout the relational algebra expression, which could be a subset of fields returned if the query is executed on the correct table schema. The missing fields are addressed by \textsc{Field} procedure in Procedure~3.

To ensure that the response generated using data synthesis can be reproduced in the web application with real database settings, we additionally extract the constraints from the query and generate a consistent result that conforms to these constraints. For this, we define \textsc{Constraints} function, which outputs all the row-based and column-based constraints in the relational algebra expression for the SMT solver. We utilize a subset of translation rules proposed by Veanes et al.~\cite{veanes2009qex}. Note that this function also assigns types to fields if the field is compared with a concrete value in the select operator or returned by set functions like \textit{COUNT}.

We provide an example for the query in Line 5 of Figure~\ref{code: php db} as input. The relational algebra expression for the query is $\text{SELECT}_{\text{A.c = 1}}(\text{A})$. Since there is no \textit{LIMIT} operation inside the query, the \textsc{MaxRow} outputs that the maximum line is infinite. The \textsc{Fields} function produces a set with one pair: $\{ \langle \text{A}, \text{c}\rangle \}$. The \textsc{Constraints} function translates the condition in the select operator to the SMT formula: \verb|(= A#c 1)| and assigns $\langle\text{A}, \text{c}\rangle$ to be of integer type.

\begin{figure} [t] \small
    \centering

\begin{lstlisting}
<?php
$conn = mysqli_connect(...);

$res = $conn->query(
    "SELECT * FROM A WHERE A.c = 1"
);

$x = $res->fetch_assoc();

$a = $x["a"];

if ($a == 0) echo 1;
\end{lstlisting}
    \caption{Example of PHP Application Database Call}
    \label{code: php db}
\end{figure}

The crucial procedures for the generation workflow are presented in Procedure~3. Before fuzzing starts in \textsc{CorbFuzzTestOne} (Line~5, Procedure~2), the \textsc{Initialization} procedure is executed. This procedure initiates three global hashtables for caching. These are preserved throughout the runtime lifecycle and synchronized throughout all runtimes 
(since we use multi-threaded distributed fuzzing, this is necessary).  

When a query is sent to the database, and the web application is waiting for the response, \textsc{Add} procedure replaces the original code for sending the query and receiving the response from the database. 
\textsc{Add} procedure takes two arguments: the query and the seed.
If the cache contains the previous solution for the query and the seed, the cached result is returned. Otherwise, the query is parsed into relational algebra to extract constraints, fields, and maximum length (as mentioned before), and an empty hashtable is returned. The hashtable, regardless of whether there is a cache hit, is tracked and used by the web application as the output of the database query.


\begin{algorithm}[] 
\label{proc:sql-imp}
\begin{algorithmic}[1]

\Procedure{Initialization}{}
  \State ConcreteResults $\leftarrow$ HashTable()
  \State Types $\leftarrow$ HashTable()
  \State Cache $\leftarrow$ HashTable()
\EndProcedure

\Procedure{Add}{Query, Seed}
   \If{Cache(Query) = NULL}
        \State ra $\leftarrow$ Parse(Query)
       \State Cache(Query).L $\leftarrow$ \textsc{MaxRow}(ra)
        \State Cache(Query).F $\leftarrow$ \textsc{Fields}(ra)
        \State Cache(Query).C $\leftarrow$ \textsc{Constraints}(ra)

    \EndIf
  \State results $\leftarrow$ ConcreteResults(Query, Seed)
  \State \Return Tracked(results)
\EndProcedure

\Procedure{Field}{Query, Seed, Name}
   \State cache $\leftarrow$ Cache(Query)
   \State r $\leftarrow$ ConcreteResults(Query, Seed)
   \If{Name $\notin$ cache.F}
        \State Types(Query, Name) $\leftarrow$ $\tau$.AssignWeight(0)
    \EndIf
    \State r.F $\leftarrow$ cache.F $\overset{+}{\leftarrow}$ Name
    \For{field $\in$ r.F}
        \If{field.Type = NULL}
          \State $\tau \leftarrow$ Types(Query, field)
          \State field.Type $\leftarrow$ Sample($\tau$)
         \EndIf
    \EndFor
    \State len $\leftarrow$ Seed \% cache.L
    \State r $\leftarrow$ Solve(cache.C $\cup \neg$ cache.Solved(r.F, len), len)
    \If{r = UNSAT} \Return Abort()
    \EndIf
    \State cache.Solved(r.F, len) $\overset{+}{\leftarrow}$ r
    \State \Return Tracked(r(Name))
\EndProcedure

\Procedure{Notify}{Query, Name, IType}
   \State $\tau \leftarrow$ Type(Query)(Name)
   \State $\tau$(IType) $\leftarrow$ $\tau$(IType) + Weight
\EndProcedure
\end{algorithmic}
\caption{Database Query Result Generation Algorithm}
\end{algorithm}

If the tracked hashtable is searched in the later executions of the web application and the searched key corresponds to NULL value, the \textsc{Field} procedure is called. In addition to the query and the seed, this procedure takes an additional argument: the name of the field (i.e., the key of the hashtable that the application is searching for). \textsc{Field} procedure assumes that the web application code is correct and the queried field must exist. Under the circumstance that this specific field name is not inferred from the SQL query statement (e.g., a wildcard select), \textsc{Field} appends the field name to the global cache so that in the future, for this query, this specific field would be considered. Before solving the constraints generated from the evaluation of the query, \textsc{Field} first probabilistically selects a type from all possible data types for each field, which is discussed in Section B. The number of rows is generated using the seed value. To avoid generating an identical result, \textsc{Field} appends constraints stating that the result to be solved should not be equal to previously generated results under the same condition (i.e., same type and same amount of row). If the solver concludes these constraints could derive no result (i.e., UNSAT), the web application immediately returns an internal error to abort the data synthesis workflow. However, this case rarely happens in our experiments because constraints for SQL queries are very permissive. The returned value of \textsc{Field} procedure is also tracked for type inference purposes, which is described in the following section. 

We demonstrate an example for the workflow over the PHP application code listed in Figure~\ref{code: php db}. Before the execution of any code, as soon as the runtime starts, the \textsc{Initialization} procedure is called. Then, on Line 2, the code calls \verb|mysqli_connect| to establish a connection to MySQL database. Inside the runtime, this function is replaced with a dummy method that always acts as if there is a successful connection. Then, the code is executed to send a query to MySQL database (Line 4) and wait for the response (Line 8). Instead of sending the query, the runtime calls the \textsc{Add} procedure. Suppose we are using a new seed, the procedure would evaluate the query and return a traced empty hashtable. On Line 10, the hashtable is searched with a key \verb|a|. Since the hashtable is empty, the key points to NULL value. Instead of returning NULL, the \textsc{Field} procedure is called to solve for all the fields, including the field searched by the application. 

\subsection{Type Inference}


The knowledge of field names is not enough to generate the data. Correct type of each field is also required for generating a consistent result. Note that for types here, we are not referring to the actual type of a concrete value. Instead, we are referring to the inherent types. The inherent type is the same after type juggling. Suppose an integer is cast as string in the application, we do not record this as string but instead as integer. Indeed, all fields in the result from the call \verb|mysqli_query| are cast as string, regardless of what the type is attributed to each of them in the table schema. Yet, they are directly used as their inherent type throughout the execution in web applications, which is made possible by type juggling. Hence, for data synthesis purposes, we need to infer the inherent types but not the actual types.

We consider type information crucial because an inaccurate type makes web applications prone to producing errors and unrealistic responses. For instance, deserializing an integer or integer-like object would inevitably lead to errors. Another example is that using a string as an index for an integer-indexed array does not lead to error but breaks the original logic of the web application. This situation is unwanted in this context because it produces a spurious response that is not reproducible in an actual run of the web application using the real table schema. 


In the query, we could gain type information for fields when the operations processing or generating the field are known and the argument or return types are well-defined. This is because type juggling in SQL would lead to an error or warning. For instance, a comparison between a field and an integer would help us conclude that the field type must be integer. However, it is impossible to infer all types from evaluating queries. Thus, we additionally infer the type of fields by the information during the execution. Specifically, data synthesis runtime collects type information via two methods. First, if the field encounters the binary comparison operand, CorbFuzz records the type of the concrete value it is comparing to. Second, CorbFuzz tracks the internal functions that the field is served as an argument. Internal functions typically have a clear definition of the types of each argument. For simplicity, CorbFuzz ignores corner cases like comparison between two fields and passing to an internal function supporting all types. Future work may leverage Hindley-Milner algorithm~\cite{milner_theory_1978} to construct a more fine-grained typing system.

Still, the runtime analysis is not enough for inferring types of all fields. Some of them may not be passed to an internal function or used in comparisons. Additionally, comparison between variables of different types is allowed, and it is impossible to deduce the inherent type of a concrete value. These factors mean there is a possibility that a different type is used against the compared variable. To accommodate these cases, we define a domain of types ($\tau$) for each field and assign a weight to any type $t \in \tau$. At initialization, each $t$ is set with an initial weight and increased whenever it matches inference (e.g., passed to an internal function), which we refer to as a type hint, after the generation of the result. If query analysis has already assigned a type, then the type would have infinite weight in $\tau$. Before constraint solving is initiated, CorbFuzz conducts a probabilistic sampling from $\tau$ for each field based on weights assigned to types (the probability of a type to be chosen is proportional to the weight of the type). Due to probabilistic selection, a variety of types are explored during fuzzing. Here, we assume that if a type for a variable is not intended, then this incorrect type used would lead to either errors or no effect on analysis. 
In general, CorbFuzz tries to increase the likelihood that a correct type will be used.

In Procedure~3, \textsc{Field} procedure conducts a probabilistic sampling over the $\tau$ for each field (Line 19-22). In our implementation, we utilize A-res algorithm \cite{vitter_random_1985}. \textsc{Notify} procedure is called when the tracked value returned by \textsc{Field} procedure is used in internal functions or for comparison. The type hint is then used to increase the weight for that type in $\tau$. In our implementation, we only let $\tau$ include integers, strings, and booleans. Type hints for types that are not in $\tau$ are ignored.  

In the example provided in Figure~\ref{code: php db} Line 10, after the \textsc{Field} procedure ends, \verb|$a| is assigned the generated value that is tracked. On Line 12, the tracked value is compared to an integer. The \textsc{Notify} procedure is called, adding weight for the integer type for the field \verb|a| in global hashtable \verb|Types|.

\subsection{Authentication Bypass Workflow}

Cookies and sessions are commonly leveraged by web applications to make HTTP requests stateful~\cite{authrev, web_two_decades}, allowing for the implementation of authentication. Both of them could be represented as a hashtable. We observed that there could be a significant increase in coverage for a web application if cookies or sessions are properly set (e.g., an authentication token presents for a specific field). It is because complex logic inside web applications tends to be reached after the request presents to be authenticated or authorized.
Usually, cookies or sessions keys and values are compared to a constant or a result from the database. Therefore, using a fuzzer to explore cookies and sessions is largely ineffective since there is a huge search space for the keys and values.

To better explore behaviors of web applications, we generate decisions for comparison operations that involve sessions or cookies. Still, we conduct concolic execution and record the constraints for the decisions made to check whether all decisions made are satisfiable. CorbFuzz treats each item in session or cookie as a pair of symbolic variables: $\langle \phi, \alpha \rangle$, where $\phi$ is the gated boolean symbolic variable that shows whether the item is defined and $\alpha$ represents the symbolic variable for the item. This method is inspired by hybrid fuzzing but differs from it. The runtime only solves the constraint one time when necessary. That is, if an item of cookies or sessions has not been passed to an operation that does not have an SMT formula translation available, the value would never be generated. 

The reason we do not generate the data as soon as it is used is largely due to the use cases of sessions and cookies. They are used in multiple or nested branches, but most of the time, their concrete value would not be evaluated. Additionally, there are very few internal functions that commonly use sessions or cookies as arguments. We have implemented only basic arithmetic and \verb|isset|~\cite{isset} internal call with the translation of the SMT formula. Still, most requests in our experiment do not require generating the concrete value of sessions and cookies. 

\begin{figure} [t] \small
    \centering

\begin{lstlisting}
<?php
session_start();

if (isset($_SESSION["is_auth"]))
    echo $_SESSION["welcome_message"];

\end{lstlisting}
    \caption{Example of PHP Application Session Usage}
    \label{code: php session}

\end{figure}

\label{proc:session-imp}
\begin{algorithm} 
    \caption{Session Generation Algorithm}
    \begin{algorithmic}[1]
        \Procedure{Initialization}{}
            \State GC $\leftarrow$ HashTable() 
        \EndProcedure
        \Procedure{Start}{}
            \State RCache $\leftarrow$ HashTable()
            \State NewSeed $\leftarrow$ Copy(Seed)
        \EndProcedure
        \Procedure{Do}{Name, Opline}
            \If{GC(Seed, Name)} 
                \State Session(Name) $\leftarrow$ GC(Seed, Name)
                \State \Return Next()
            \EndIf
            \If{Opline.Operand $\in$ ImplementedOp}
              \State decision $\leftarrow$ NewSeed \& 1 
              \State ShiftRight(NewSeed)
              \State cons $\leftarrow$ ToConstraint(Opline, decision)
              \If{$\neg$ IsSAT(RCache(Name) $\cup$ cons)}
                \State \Return \textsc{Do}(Name, Opline)
              \EndIf
              \State RCache(Name) $\overset{+}{\leftarrow}$ cons
              \State \Return decision
            \Else
              \State solved $\leftarrow$ Subset(GC(*, Name))
              \State cons $\leftarrow$ RCache(Name) $\cup \neg$ solved
              \If{$\neg$ IsSAT(cons)} \Return Abort() 
              \EndIf
              \State GC(Seed, Name) $\leftarrow$ Solve(cons)
              \State \Return \textsc{Do}(Name, Opline)
            \EndIf
        \EndProcedure
    \end{algorithmic}
\end{algorithm}

We have shown the crucial components for the workflow in Procedure~4 for session, which is identical for cookie. Similar to the previous workflow for database, there is also an initialization procedure that creates a global hashtable for caching. Specifically, \verb|GC| is for storing the mapping between seed and the sessions. Additionally, there is a \textsc{Start} procedure, which is called before each HTTP request is handled and the variable declared only survives during the lifecycle of that request. The procedure creates a copy of the seed and declares a hashtable for saving the constraints for each session item used during the request.

When an item of sessions is compared with a concrete value, the \textsc{Do} procedure is used before evocation of the original comparison handler. CorbFuzz first checks whether there is already a cached item for the given seed (Line~7). If there is a cache hit, then the item is assigned a concrete value, and the internal implementation of the comparison operation is executed. Otherwise, CorbFuzz checks whether the comparison operation is implemented (Line~10) so that it could convert the decision of the operation to a constraint. If so, a decision is generated from the seed, and the constraint for performing this decision is appended to the constraints over that item (Lines~11-13). An SMT solver is then used to check whether the constraint is satisfiable (Line~14). If it is not satisfiable, then the procedure recursively consumes the seed until there is a decision that could be satisfied. Our implementation assumes there are at most 32 decisions since we are using a 32-bit seed.  In our experiments, the maximum consumption is only 11 bits in a specific request. The decision is then returned, and the internal implementation of the comparison is ignored. As for the corner case that a session item is compared to another session item, we treat this comparison as an unimplemented operation for one side and then apply the workflow to the other side. When the operation is not implemented, then a concrete value is generated by solving the constraint for that item. To ensure the uniqueness of the concrete value generated, the solver tries to avoid generating already solved values stored in the global cache for that field name. To reduce UNSAT cases, \textsc{Do} only selects a random subset of the stored values in the cache and removes them from consideration as the result of solving (Line~19). Note that by doing so, we do not create a strict bijection here between the seed and the sessions. Same sessions may map to multiple seeds. This is because the constraints here are not permissive. 

For PHP code listed in Figure~\ref{code: php session}, when it executes until Line~4, CorbFuzz first declares a pair of symbolic variables $\langle\phi_{0}, \alpha_{0}\rangle$ and makes a decision for the unary comparison \verb|isset| based on the seed. Suppose the seed indicates the decision is to return true, then the constraint $\phi_0 = \text{true}$ is added to the set of constraints for the \verb|$_SESSION["is_auth"]|. Note that this session item is not used later, so its concrete value is never generated. Then, on Line 5, another session item is used. We have not implemented \verb|echo| function and the value of \verb|$_SESSION["welcome_message"]| is generated with respect to its constraints (i.e., no constraint in this context).

In certain cases, the data stored in the cookies or sessions may be subject to decryption or deserialization in web applications. Before a decryption or decoding function is executed with input from cookies or sessions, the workflow must synthesize the concrete value. This situation is undesirable because authentication cannot be bypassed. A more general version of this issue, which is that symbolic execution fails to model a comprehensive list of syscalls, also plagues hybrid fuzzing\cite{Driller}, and no solution has been proposed so far. We discuss a potential ad-hoc solution and future work to address this problem in Section~\ref{subsec:eval-data-synthesis}.


\subsection{Adapting to Other Programming Languages}

The aforementioned methods target PHP applications. However, they can be extended to other programming languages that are widely used for web application development. Specifically, the data synthesis workflows embedded with type inference can be applied to other programming languages supporting type juggling, like Perl or JavaScript/NodeJS, or using a dynamic type system, like Ruby and Python. For statically typed programming languages, like Golang or Java, the type inference component would not be needed, but the data synthesis workflows can similarly be adapted. 


\section{Implementation \& Evaluation}
\label{sec:evaluation}
We have implemented the coverage-guided fuzzer and oracle for CorbFuzz in Python with 900 lines of code (LoC) for fuzzing web applications written in PHP. Unlike existing web application fuzzers that only consider responses related to PHP, CorbFuzz considers all responses after a web page is loaded, including images, CSS, and RPC communications. The data synthesis workflow is implemented as an external module with 500 LoC in C and 1200 LoC in NodeJs for PHP. PHP 7.4 has been instrumented to support the workflow and provide branching information for coverage evaluation. To allow for fair evaluation on data synthesis effectiveness, we implement two baseline workflows by removing components inside CorbFuzz. 

In the following subsections, we address the following research questions;

\noindent
\textbf{RQ1.} Is data synthesis workflow generating consistent data?

\noindent
\textbf{RQ2.} Can data synthesis workflow increase test coverage?

\noindent
\textbf{RQ3.} Can CorbFuzz detect bugs in implementation in existing browsers?

\subsection{Experimental Setup}
\subsubsection{Environment} We evaluate CorbFuzz on two Intel Xeon Phi 7210 (64 cores) nodes. Both nodes use Ubuntu 20.04 with one node running NGINX\cite{noauthor_nginx_nodate} for serving web application on the instrumented PHP environment and other node running the coverage-guided fuzzer. 

\subsubsection{Targets} We evaluate CorbFuzz with two popular web browsers: Chromium and WebKit (Safari). Chromium has already added CORB into its current stable release. We implement a test harness based on the Chromium shared library containing the CORB implementation. For WebKit, the developers have created a pull request for CORB implementation but it has not yet been merged into the main branch. Since its implementation is relatively simple and straightforward, we directly translate it into Python to implement a test harness for CORB implementation of Webkit. 

\subsubsection{Web Applications} 

\begin{table}[]
\begin{tabular}{|l|r|r|}
\hline
{\bf LoC Range} & {\bf Number of Applications} & {\bf Average LoC} \\ \hline
Less than 1K 
& 15                & 476.9       \\ \hline
Between 1K and 10K 
               & 15                & 3022.5      \\ \hline
Between 10K and 100K            & 6                 & 43075.5     \\ \hline
More than 100K   & 3                 & 250875.5    \\ \hline
\end{tabular}
\centering
\caption{Total LoC Statistics for fuzzing targets}
\label{table: loc}
\end{table}

Web applications are fuzzed to provide responses as input for browser policy test harnesses. We crawled 300 repositories containing PHP code from GitHub between March 2nd, 2021 and April 10th, 2021. The repositories are filtered out if they do not contain \verb|index.php| or \verb|index.html|. For simplicity, we do not consider applications that require downloading dependencies with Composer\cite{noauthor_composer_nodate}, a dependency management tool.  
The number of remaining applications is 58 with varying LoCs. We fuzz the policies with these 58 applications but for the sake of evaluation of data synthesis effectiveness, we only use 39 of them, for which CorbFuzz reports existence of branches and utilization of databases. The statistics of these applications are presented in Table~\ref{table: loc}.

\begin{figure} [t] \small
\hspace*{-9mm}
    \centering
    \includegraphics[width = 105mm]{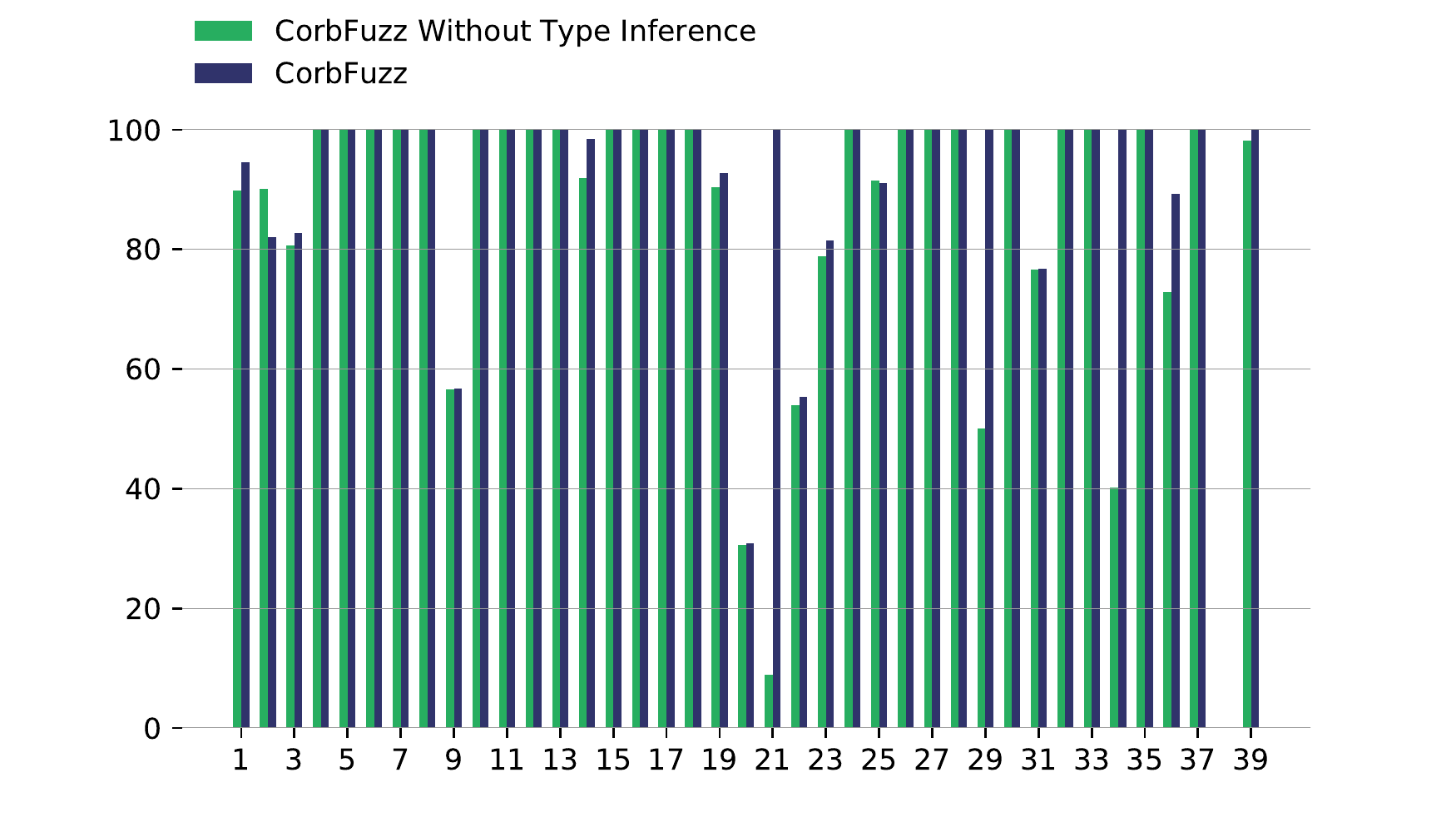}
    \caption{The percentage of correct type generation for comparison statements for CorbFuzz and CorbFuzz without Type Inference for 3 minutes of fuzzing. X axis denotes the web application ID.}
    \label{fig:type-gen-comp}
\end{figure}

\begin{figure} [t] \small
\hspace*{-9mm}
    \centering
    \includegraphics[width = 105mm]{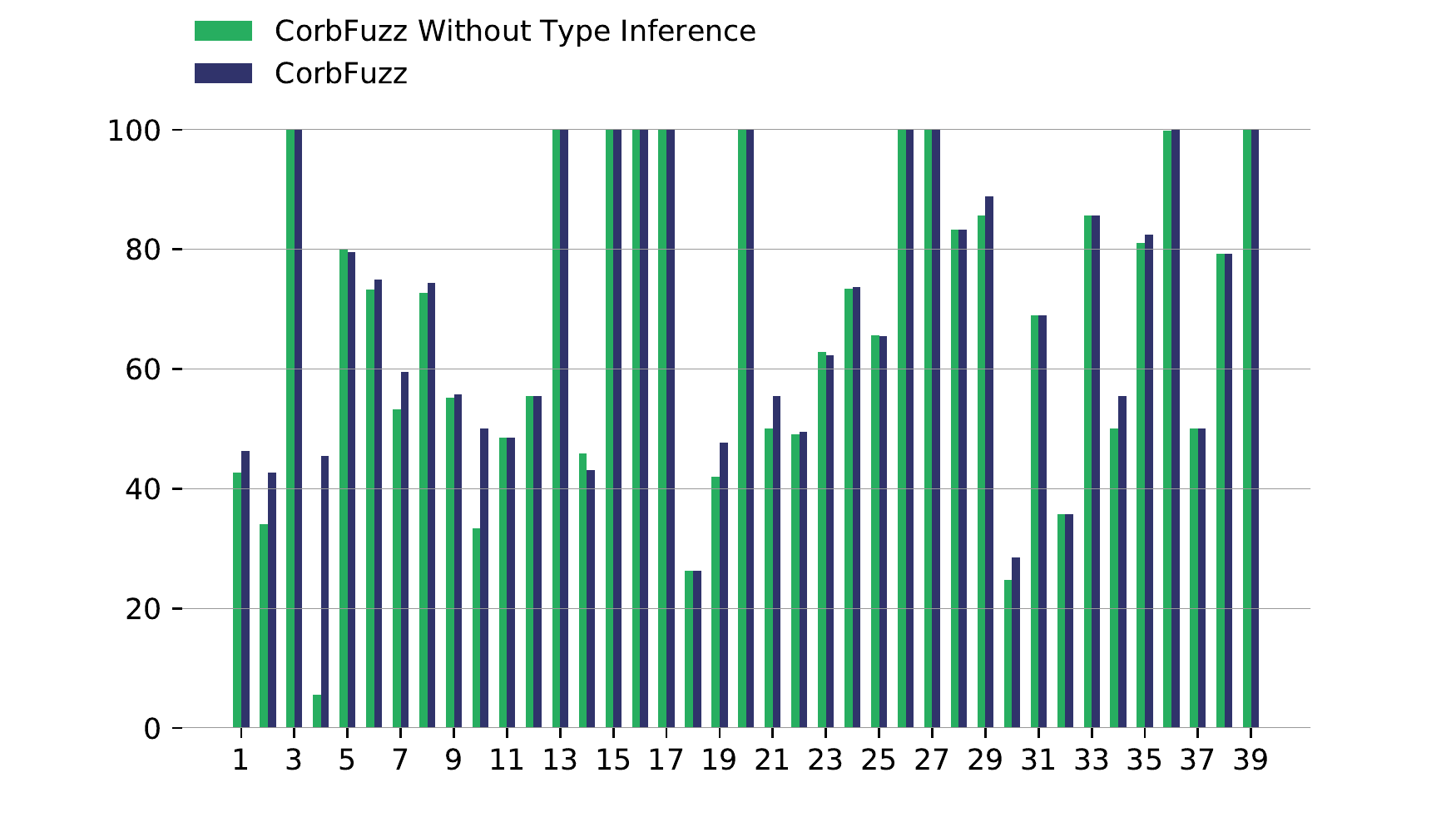}
    \caption{The percentage of correct type generation for internal function calls for CorbFuzz and CorbFuzz without Type Inference for 3 minutes of fuzzing. X axis denotes the web application ID.}
    \label{fig:type-gen-fn}

\end{figure}

\subsection{Data Synthesis Effectiveness}
\label{subsec:eval-data-synthesis}

To evaluate the data synthesis approach and address \textbf{RQ1}, we ran CorbFuzz with and without type inference for three minutes\footnote{Due to the randomness feature of the fuzzer, we ran each experiment five times and take maximum value (e.g., edge coverage).} with each web application. We compared the percentage of correct type generations by tracking whether the generated value matches the type of (1) a concrete value when compared to it; (2) the defined argument when used to call internal function. Figures~\ref{fig:type-gen-comp} and~\ref{fig:type-gen-fn} demonstrate the percentage of correct generations for comparison statements and internal function calls respectively. Figure~\ref{fig:type-gen-comp} shows that for 10 applications, CorbFuzz generates the correct type for comparisons more often than CorbFuzz without type inference with 17\% more data generations with correct type on average. Figure~\ref{fig:type-gen-fn} shows that for 11 applications, CorbFuzz generates the correct type for internal function calls more often than CorbFuzz without type inference with 5\% more data generations with correct type on average. 

On some applications, CorbFuzz has little improvement on the accuracy of type generation because in the results we show, we consider all type violations. However, many of these type violations are due to developers using type juggling and not due to data synthesis. Therefore, these violations cannot be removed by improving type inference in data synthesis. 



\begin{figure} [h] \small
\hspace*{-9mm}
    \centering
    \includegraphics[width = 105mm]{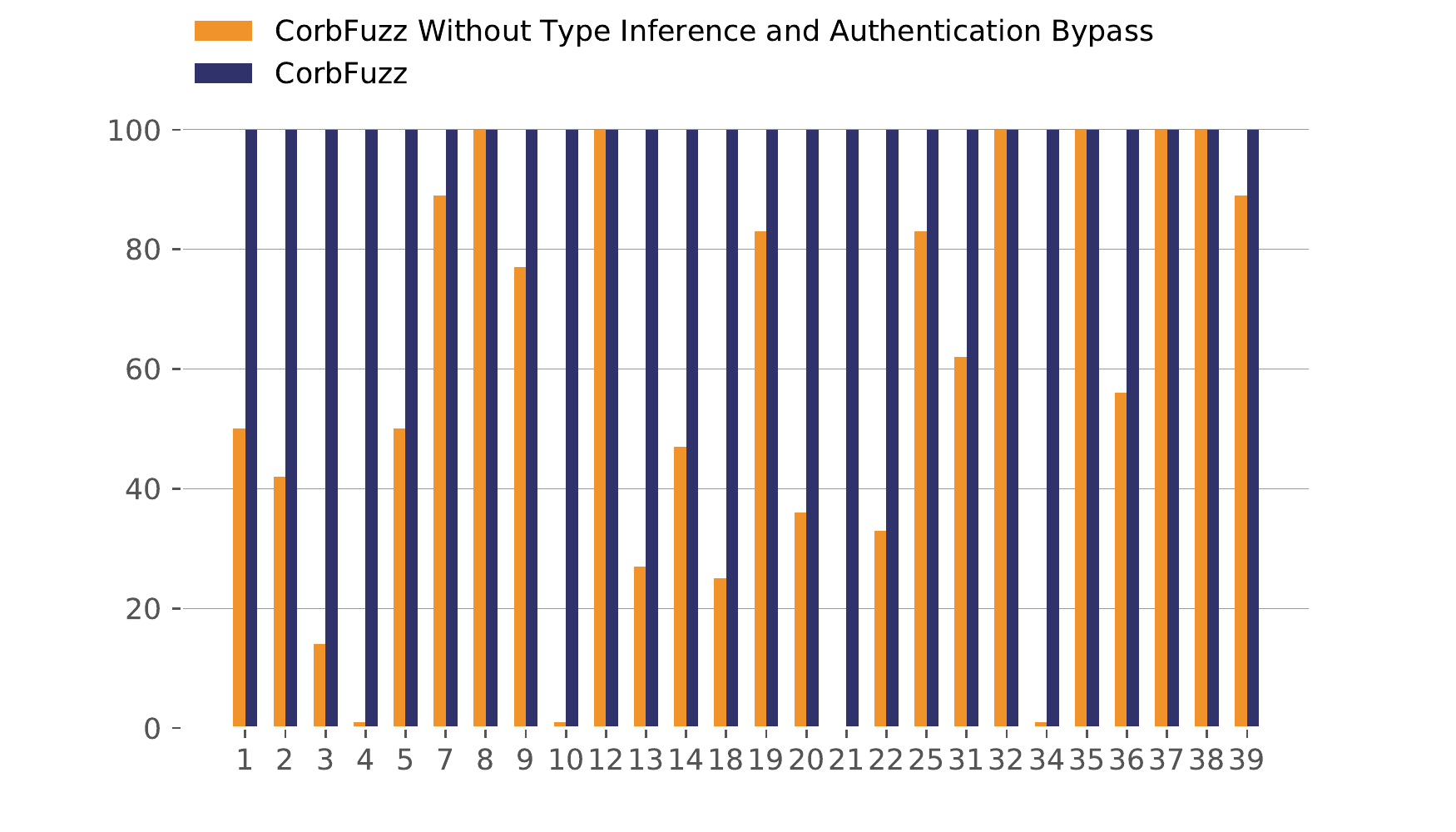}
    \caption{The percentage of edges covered for CorbFuzz without Type Inference and Authentication Bypass in 3 minutes of fuzzing against edges covered by CorbFuzz. X axis denotes the web application ID.}
    \label{fig:edge-cover-perc}

\end{figure}


We also evaluated the impact of data synthesis on fuzzing effectiveness and address \textbf{RQ2}. Figure~\ref{fig:edge-cover-perc} demonstrates the edge coverage difference, in terms of percentages, of edge counts between CorbFuzz without any type inference and authentication bypass, and CorbFuzz. For this evaluation, we only chose applications containing more than one branch. Figure~\ref{fig:edge-cover-perc} shows that for almost all applications, the inclusion of type inference and authentication bypass improves coverage. The average number of edges covered is 16.2 edges for CorbFuzz without type inference and authentication bypass and 27.5 for CorbFuzz. These results demonstrate that with the inclusion of type inference and authentication bypass, we can cover on average 70\% more edges, which shows the effectiveness of our data synthesis approach.

The number of edges covered is low for some applications because these applications (e.g., WordPress) save and use structural, encrypted, or serialized data from the database or cookie. The data synthesis workflow is unaware of the structural property of any field. Therefore, it generates a large amount of data that can not be deserialized or decrypted by the web application, so CorbFuzz fails to explore these web applications. However, we recognize that this can be prevented by enlarging the domain of type $\tau$ defined. By considering the common structural properties as types (e.g., JSON type) and instrumenting deserialization libraries to provide type hints, future work could implement an approach that is able to further improve web application coverage.

\subsection{Detected CORB Weaknesses}

In evaluation of {\bf RQ3}, we have discovered three common classes of code patterns, which are discussed in following sections, that cause the CORB implementation in Chromium to not function as expected. One of the cases has been filed in the Chromium bug tracker before our discovery by a Chromium developer and is still in discussion\footnote{https://crbug.com/795470}. We have reported another {case}\footnote{https://crbug.com/1148397}, which has later been resolved by a patch in the CORB component {of Chromium}\footnote{https://chromium-review.googlesource.com/c/chromium/src/+/2596879/}.

All the weaknesses discovered are due to the novel fuzzing approach we present in this paper. The weaknesses are not present in every web application but only in a few of them. Fuzzing a single web application would likely not lead to the discovery of any of the weaknesses, while fuzzing multiple web applications without the data synthesis would require setting up the database and seeding the tables for numerous web applications, which is infeasible. Data synthesis allows us to effortlessly fuzz multiple web applications by symbolically evaluating database queries. Additionally, all the web application states that lead to the aforementioned weaknesses are under some extent of authentication or authorization. Without a manual definition of the login method for a web application, the fuzzer would not discover these weaknesses. Yet, manually defining authentication or authorization method for a considerable number of applications is tedious and unrealistic. In contrast to the manual approach, the data synthesis approach we present automatically generates appropriate sessions and cookies items, which allows exploration using authorized requests.

\begin{figure} [t] \small
    \centering
    \includegraphics[width = 80mm]{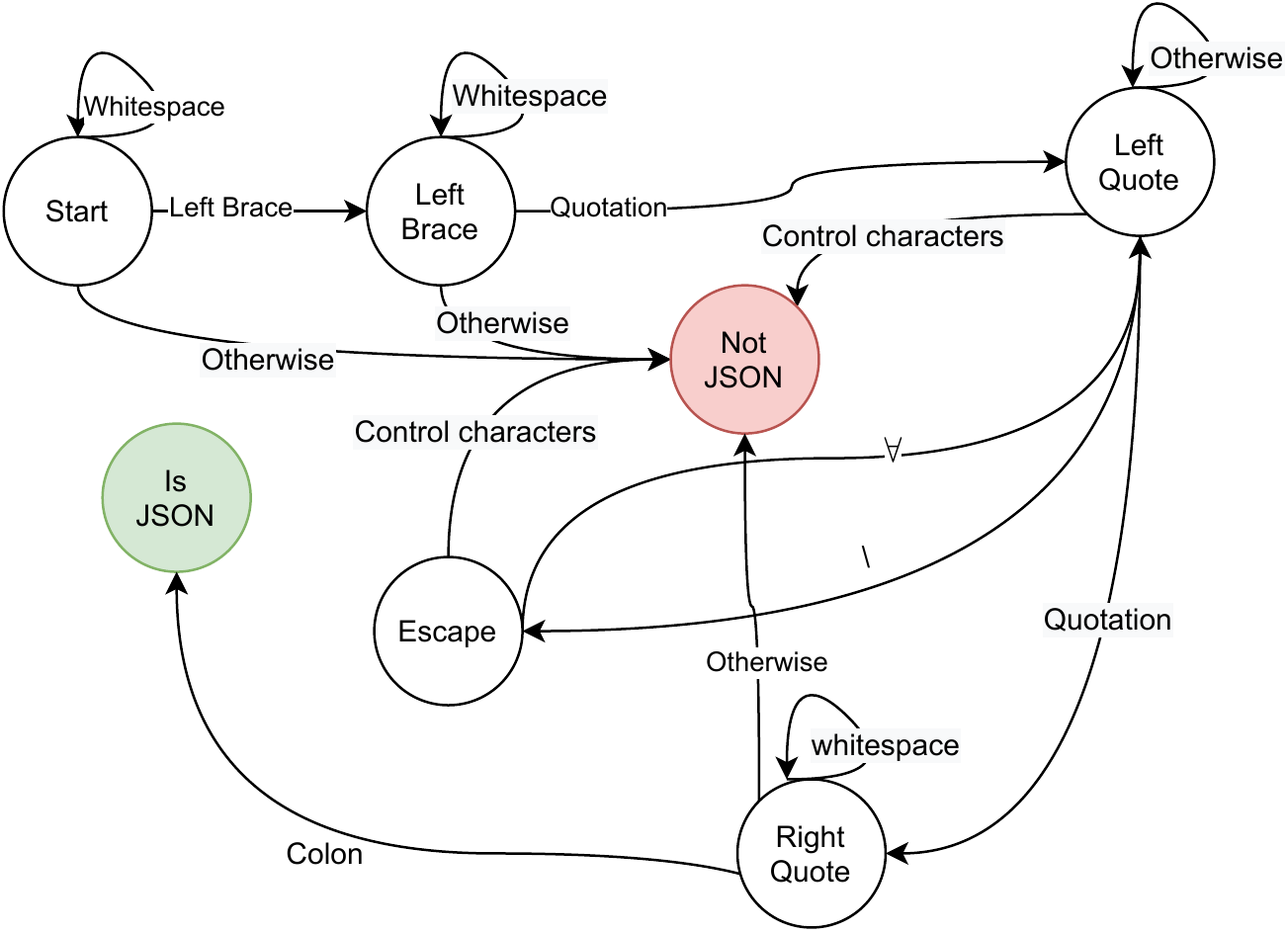}
    \caption{Finite State Machine for Validating JSON}
    \label{fig: json fsm}

\end{figure}

\textbf{Serialized Array as JSON Response.} In Chromium, if the response MIME type is related to JSON, CORB would check the response content to learn whether it is indeed JSON. A finite state machine (FSM) conducts such a check. As illustrated in Figure~\ref{fig: json fsm}, the FSM does not comprehensively parse the response content to perform the check. Instead, it only checks whether the content has a left brace at the beginning and has matching quotes for the first key to determine if the content is JSON.

As permissive as it is, such a check would not identify a serialized array in JSON format, which is considered a JSON object inside the JavaScript runtime of Chromium. Indeed, the latest JSON specification (RFC 8259\cite{ietf_tools}) refers this to be a different type from JSON object known as JSON array. For instance, for a simple response as \verb|[1,2,3]|, a JSON array, JSON check in CORB implementation would first look for the left brace. Yet, the first character is left bracket, which makes the FSM classify the content as not JSON. 
However, \verb|fetch|, \verb|XMLDocument|, and \verb|JSON.parse| APIs in JavaScript runtime  parse the content into a JavaScript object without warning.

Sending JSON arrays as responses is commonly seen in web application APIs. The responses of these APIs would likely carry sensitive information. Thus, we consider catching JSON array for JSON MIME type in CORB implementation to be a reasonable patch.

\textbf{Malformed JSON Response.} It is not uncommon for web application developer to adopt the following code pattern, where \verb|$var| represents any variable the attacker can control (i.e., a tainted variable), which could be achieved through methods including URL manipulation and security-unrelated CSRF\cite{shankar_d1r3wolf_chaining_nodate}.

\begin{lstlisting}
<?php
header('Content-Type: application/json');
echo "{\"$var\":\"$secret\"}";
\end{lstlisting}

This code pattern does not leverage the existing serialization library. Instead, it produces serialized objects by direct string concatenation and manipulation. If the attacker is able to control at least one character in the first key of a JSON object, they would be able to bypass the CORB check by making that key contain a control character. According to JSON specification, control characters (U+0000 through U+001F) inside key and value of JSON object should be escaped (i.e., append a reverse solidus before the control character). Similarly, the JSON verifier inside CORB implementation in Chromium follows this pattern and rejects all JSON objects with unescaped control characters on the first key. 

Consider the PHP code shown above. If we set \verb|$var| to be the control character \verb|\u0017|, the resulting response becomes \verb|{"\u0017": "[SECRET]",[MORE SECRETS]}|. The JSON checker FSM enters the state ``Left Quote'' after encountering the first and second characters. It then compares character \verb|\u0017| to control character range and identifies it as an unescaped control character, misclassifying the response as not JSON.


We consider this weakness should be addressed as the existence of such a code pattern is not negligible. We have reported this to the Chromium team, and it has been fixed by removing the check for control character inside the JSON checker. 

The discovery of this weakness is only possible if the fuzzer can mutate the database state efficiently. In all the cases that lead to this malformed JSON response, the variable as the key of the JSON array is retrieved from the database. CorbFuzz symbolically evaluates the database queries and synthesizes the concrete value for that variable, which allows the efficient exploration of the domain of this variable (i.e., exploring interesting values for UTF-8 character). In contrast, a conventional fuzzing approach has to mutate the database state by sending requests to the web application or by resetting the database, leading to significantly larger search spaces and high timing overhead for mutating the content of the variable derived from the database.


\textbf{Confirmation Sniffing.} In most web applications, warnings and errors in plaintext or HTML are directly prepended to the response. For PHP, a warning in HTML is generated whenever an undefined behavior happens. If a malicious actor is able to trigger an undefined behavior in responses that are checked with confirmation sniffing, then CORB in Chromium could be bypassed since the responses start with data that is not of their MIME type.

This weakness, including all previous weaknesses, could be considered as the side effect of increase in permissiveness caused by confirmation sniffing. We consider that confirmation sniffing is harming the effectiveness of the CORB implementation in Chromium. Future work, on the other hand, could work on testing the contribution of confirmation sniffing on compatibility and conclude whether confirmation sniffing is indeed redundant. 

\subsection{Fuzzer Flexibility}
We have constructed an oracle for ORB and test the proposed implementation. Our fuzzer is unable to discover any weakness of ORB. It is because ORB applies a whitelist approach to block requests yet CORB uses a blacklist, which means ORB is much less permissive than CORB. Future work could apply similar approach to evaluate its compatibility.

\section{Related Work}
\label{sec:related-work}

\textbf{Coverage-guided Fuzzing.}
Coverage-guided fuzzing has been used to find bugs in different types of programs, such as virtual machines~\cite{jvmfuzz}, web browsers~\cite{xu_freedom:_2020,noauthor_fuzzilli_2021}, network functions\cite{p4rl, porkfuzz}, and operating systems\cite{kim_finding_2019,kim_hfl:_2020}. The state-of-the-art implementations are AFL~\cite{AFL} and libFuzzer~\cite{libfuzzer}.
In this paper, we leveraged coverage-guided fuzzing to explore responses from web applications for browser security policy checking. Yet, our approach is not using the coverage of the browser but is instead guided by the coverage of the web applications. Our approach also conducts a series of coverage-guided fuzzing with different targets instead of fuzzing an individual program.


\textbf{Browser Fuzzing.} 
Domato~\cite{noauthor_domato_2021}, Dharma~\cite{noauthor_dharma_2021}, and FreeDom~\cite{xu_freedom:_2020} are all specialized fuzzers used to discover memory-related vulnerabilities and assertion violations in DOM implementation of browsers. They generate structural data that contain valid HTML, CSS, and DOM-related JavaScript for browsers to render. Fuzzilli\cite{noauthor_fuzzilli_2021} and Jsfunfuzz\cite{noauthor_collection_2021} are fuzzers for discovering vulnerabilities in JavaScript engines, which utilize a similar approach to generating structural data. Our work is different from all these approaches since the oracle of CorbFuzz is defined based on the property of the web applications, and CorbFuzz does not generate the test cases but instead utilizes web applications' responses. 
Roy et al.~\cite{roy2014x} fuzzes web applications and supplies responses to browsers to detect visual inconsistencies between browsers. It is similar to our work in the sense that both works treat web applications and browsers together as a black box. Unlike their work which focuses on testing web applications, our work focuses on testing security policies in browsers. We also do not cross-reference between browsers but use an oracle instead. 

\textbf{Web Application Testing.}
Alshahswan et al.~\cite{alshahwan2011automated} and Biagiola et al.~\cite{biagiola2017search} propose search-based approaches to testing web applications. Both works use metaheuristic approaches such as genetic algorithms to explore and generate different inputs to extensively test web applications.
Different from our work, \cite{alshahwan2011automated} requires the input types and login information. \cite{biagiola2017search} requires Page Objects to be provided to test the web application. Our work instead avoids manual analysis through data synthesis. 
Elbaum et al. \cite{sessiontest} proposes that web application testing should mutate the sessions and provides a few mutation techniques that could help achieve better coverage. Data synthesis in our work is different than the work of Elbaum et al. since we do not mutate the sessions but instead symbolically evaluate or generate them. Apollo~\cite{apollo} and Wassermann et al.~\cite{phpdym} leverage concolic testing to increase coverage of web applications and to discover vulnerabilities. Session generation workflow in our data synthesis approach is utilizing concolic execution, but it is fundamentally different than the concept of concolic testing.

%




\section{Conclusion}
\label{sec:conclusion}
We have created a browser policy fuzzer CorbFuzz which uses web application responses to fuzz the browser security policies. To avoid setting up the web applications manually, we proposed a web application runtime that synthesizes data. The resources queried by the web application are either generated or symbolically represented. We have shown that the data synthesis approach not only generates consistent data but also increases test coverage for web applications. We have evaluated CorbFuzz on CORB implementations of Chromium and WebKit as well as ORB proposal for Firefox. By fuzzing with 58 applications, we discovered three classes of weaknesses in CORB implementation of Chromium. 


\bibliographystyle{IEEEtran}
\bibliography{biblio}
\end{document}